\documentclass[sigconf,nonacm]{acmart}
\usepackage{tabularx}
\usepackage{textcomp} 
\usepackage[T2A,T1]{fontenc}
\usepackage[utf8]{inputenc}
\usepackage[russian,main=english]{babel} 

\begin{document}

\title{Russian trolls speaking Russian: \\Regional Twitter operations and MH17}

\author{Alexandr Vesselkov}
\affiliation{%
  \institution{Department of Military Technology, National Defence University}
  \streetaddress{P.O. Box 1212}
  \city{Helsinki}
  \country{Finland}}
\email{vesselkov.as@gmail.com}

\author{Benjamin Finley}
\affiliation{%
  \institution{Department of Computer Science, University of Helsinki}
  \streetaddress{}
  \city{Helsinki}
  \country{Finland}}
\email{benjamin.finley@helsinki.fi}

\author{Jouko Vankka}
\authornote{Corresponding author}
\affiliation{%
  \institution{Department of Military Technology, National Defence University}
  \streetaddress{P.O. Box 1212}
  \city{Helsinki}
  \country{Finland}
}
\email{jouko.vankka@mil.fi}

\begin{abstract}
  The role of social media in promoting media pluralism was initially viewed as wholly positive as social media could break the oligopoly of (often state-owned) mainstream media. However, some governments are allegedly manipulating social media by hiring online commentators (also known as trolls) to spread propaganda and disinformation. In particular, an alleged system of professional trolls operating both domestically and internationally exists in Russia. 
  
  To improve transparency on trolls’ influence on social media, Twitter released in 2018 longitudinal data on accounts identified as Russian trolls and their tweets, starting a wave of quantitative research on Russian trolls. However, while foreign-targeted English language operations of these trolls have received significant attention, no research has analyzed their Russian language domestic and regional-targeted activities. This is despite the fact that half of the tweets in the Twitter-released data are in Russian. We address this gap by characterizing the Russian-language operations of Russian trolls using the Twitter data. We first take a broad view with a descriptive and temporal analysis, and then focus in on the trolls’ operation related to the crash of Malaysia Airlines flight MH17, one of the deadliest incidents in the conflict in Ukraine.
  
  Among other things, we find that Russian-language trolls have run 163 hashtag campaigns (where the use of a hashtag grows abruptly within one month). The main political sentiments of such campaigns are praising Russia and Putin (29\%), criticizing Ukraine (26\%), and criticizing the United States (US) along with Obama (9\%). Further, we discovered that trolls actively reshared information. Namely, 76\% of tweets were retweets or contained a URL. Particularly often trolls distributed the news of mainstream media. Additionally, we observe periodic temporal patterns of tweet arrival, with three distinct periods that change over time, suggesting that trolls use automation tools for posting. Further, we find that trolls’ information campaign on the MH17 crash was the largest in terms of tweet count. However, around 68\% of tweets posted with MH17 hashtags were likely used simply for hashtag amplification. With these tweets excluded, about 49\% of the tweets suggested to varying levels that Ukraine was responsible for the crash, and only 13\% contained disinformation and propaganda presented as news. Interestingly, trolls promoted inconsistent alternative theories for the incident. Namely, half of the false news tweets suggested that Ukraine downed the plane with an air-to-air missile, whereas 23\% promoted the ground-to-air missile version.
\end{abstract}




\maketitle
\section{Introduction}
The advent of social media marked a significant change in media production and distribution, which was previously monopolized by large, often state-owned institutions \cite{Loader2011}. Social media enables users to share their own opinions and present alternative viewpoints, thereby changing users’ roles from passive content consumers to prosumers (producers and consumers). Social media helps build online communities, encourages debate, and mobilizes users, and therefore can promote democracy \cite{Price2013}.

However, despite the initial enthusiasm about the democracy-promoting role of social media, such media can be manipulated by various parties \cite{Arif2018}. For example, non-democratic governments allegedly hire online commentators to spread speculation, propaganda, and disinformation so as to push agendas and manipulate public opinion \cite{Sobolev2019, Zannettou2019b}. Such paid commentators are referred to as trolls \cite{Mihaylov2015, Llewellyn2019}.

An alleged system of professional paid trolls exists in Russia coordinated by a company called Internet Research Agency (IRA) \cite{OfficeoftheDirectorofNationalIntelligence2017}, which is also referred to as a troll factory (e.g., \cite{Evstatieva2018}). The Russian trolls are alleged to have interfered with several international political and social events, most prominently the 2016 US presidential election \cite{OfficeoftheDirectorofNationalIntelligence2017}. Such claims have triggered a new research stream on the operations and impact of paid Internet trolls, particularly those associated with IRA (e.g., \cite{gorrell2019partisanship, Howard2018, Zannettou2019b}). However, such studies are still scarce \cite{Tucker2018}. Furthermore, while foreign-targeted operations of Russian trolls have received some attention, as far as we know, no studies have analyzed their domestic and regional-targeted activities.

To address this gap, this paper analyzes the domestic and regional operations of Russian trolls by studying their Russian-language Twitter posts (tweets) from a dataset recently released by Twitter \cite{Twitter2019}. Although the procedure of troll detection adopted by Twitter remains unclear, we assume that the identified troll accounts indeed belonged to paid and centrally coordinated commentators. This study is the first to characterize the domestic and regional operations of Russian trolls based on the data provided by Twitter. We discover that the crash of Malaysia Airlines flight MH17 has prompted the largest information campaign of Russian-language trolls. Therefore, the paper further focuses on the trolls’ reaction to the crash of MH17.

Overall, the issue of state-sponsored organized trolls remains important with novel disinformation campaigns even targeting the COVID-19 pandemic \cite{gabrielle2020}.

This paper is structured as follows. Section \ref{data} introduces the data analyzed. Section \ref{relatedwork} presents the related work. Section \ref{content_analysis} analyzes hashtags, retweets, and URLs that Russian trolls’ used in their domestic and regional operations. Section \ref{temporal_patterns} studies temporal posting patterns of troll tweets. Section \ref{mh17} focuses on the MH17-related activities of trolls. Finally, results are discussed and conclusions are drawn in Section \ref{discussion}. 

\section{Data} \label{data}
In 2017, due to the investigation into Russian involvement in the 2016 U.S. elections and to increase transparency into foreign influence on its platform, Twitter released a list of accounts believed to be associated with the IRA. The troll tweets posted from these accounts were then first collected and released by researchers from Clemson University\footnote{https://github.com/fivethirtyeight/russian-troll-tweets} and later by Twitter itself \cite{Twitter2019}.

Specifically, between October 2018 and September 2019 Twitter released 9,691,682 tweets published between May 2009 and November 2018 by 3843 troll accounts related to the IRA \cite{Twitter2019}. The dataset includes all public, non-deleted tweets from such accounts.

The dataset contains the following information:

\begin{itemize}
\item \textbf{Tweet}: ID, language, text, time, and name of client application used to post the tweet. And when applicable ID of the tweet and user that the current tweet replied to, reposted (retweeted), quoted, or mentioned, number of times the current tweet was quoted, replied to, liked, and retweeted\footnote{The count of likes and retweets exclude the engagements from users who are suspended or deleted at the moment of data release (e.g., trolls)}, list of hashtags, URLs, and tweet geolocation.

\item \textbf{User/Profile}: ID, anonymized display and screen names, self-reported location, description of profile, language, creation date, number of followers and followed accounts at the time of suspension. 
\end{itemize}

\begin{figure}[tbp!]
\centering
\includegraphics[width=\columnwidth]{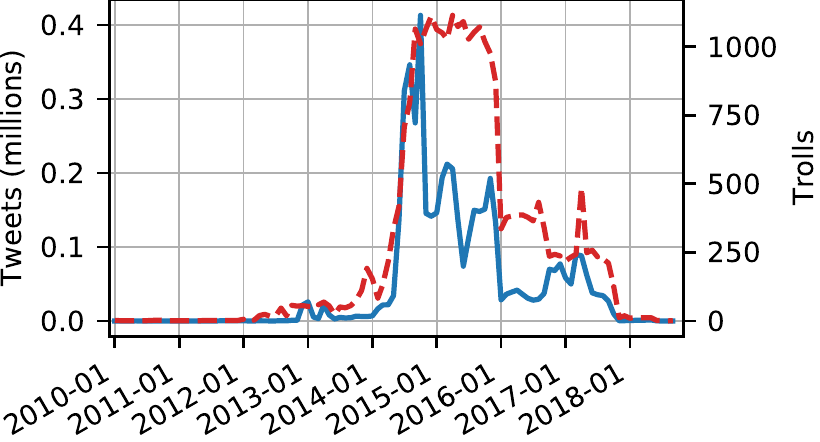}
\caption{Number of troll tweets (solid blue) and trolls (dashed red) over time}
\Description{97\% of tweets were published in 2014-2017. The number of active trolls exceeded 1000 from September 2014 to October 2015 while the number of tweets varied from 74000 to 413000 during these months.}
\label{fig:total_number_of_tweets_and_trolls}
\end{figure}

The language of 50.15\% (around 4.86M) of the troll tweets is Russian. We focus on and perform analyses on these Russian language tweets and trolls (though we also perform some comparisons against the English language trolls and tweets). These tweets were posted by 1551 accounts between January 2010 and September 2018, with 97\% of these tweets falling between 2014 and 2017. The number of trolls with at least one tweet in a month exceeded 1000 from September 2014 to October 2015 and reached the maximum of 1115 in April 2015. At the same time, the number of tweets posted during these months varied considerably from 74K to 413K. For reference, Figure \ref{fig:total_number_of_tweets_and_trolls} details the number of tweets and trolls per month over time. Further, we only analyze tweets posted in 2014-2017 to concentrate on the period of trolls' highest activity.

\section{Related Work} \label{relatedwork}
\subsection{Quantitative studies of Russian trolls}

The data on troll accounts and tweets released by Twitter has started a wave of quantitative research on Russian trolls. Table \ref{tab:quantitative_studies} provides an overview of such studies. 

\begin{table*}[tbp!]
\caption{Quantitative studies of the activity of organized trolls on social media}
\label{tab:quantitative_studies}
\begin{tabular}{p{0.5cm}p{5.5cm}p{7cm}p{3cm}}
\toprule
Ref & Data & Research objectives & Scope (case) \\ 
\midrule
 \cite{Addawood2019} & 1.2M tweets by 1148 troll accounts; 
 
 12.4M tweets by 1.2M ordinary users & \textit{Detect}: Identify linguistic markers of deception in trolls' posts and test their applicability for troll detection & US Elections; Broad, exploratory \\
 
 \cite{Badawy2019} & 540K “original” tweets by 1148 troll accounts; 13.6M tweets of ordinary users & \textit{Assess the impact}: Understand what users fall for propaganda spread by IRA trolls & US Elections \\
 
 \cite{Boyd} & 3500 Facebook ads ordered by IRA; 
 
 tweets of 969 trolls and 1078 ordinary users & \textit{Detect}: Conduct linguistic analysis to separate the posts of IRA trolls from native English-speaking users & US Elections; Broad, exploratory \\
 
 \cite{Broniatowski2018} & 899 vaccination-related troll tweets; 1.8M tweets of ordinary users & \textit{Describe}: Understand the role of bots and trolls in vaccination-related discussion & Vaccination debate
 \\

 \cite{Ghanem2019} & 1.8M tweets by 2K trolls;
 1.9M tweets by 95K ordinary users & \textit{Detect}: Define approach of detecting trolls on Twitter based on textual features & US Elections
 \\
 
 \cite{gorrell2019partisanship} & 9M tweets by 3.8K trolls;
 
 13.2M tweets by 1.8M ordinary users;  & \textit{Describe; assess the impact}: Define the role and impact of “politically-motivated actors”, including IRA trolls & Brexit; Broad, exploratory
 \\
 
 \cite{Howard2018} & Data provided by Facebook, Twitter, and Google to Senate: posts, ads, accounts & \textit{Describe}: Analyze trolls' activity over several social media platforms & US Elections and other political events
 \\
 
 \cite{Im2019} & 347K tweets by 2.2K trolls; 
 
 30M tweets by 171K ordinary users & \textit{Detect}: Develop machine learning models to detect trolls, apply the models on currently active accounts & Broad, exploratory
\\

 \cite{Kim2019} & 1.7M tweets by 733 trolls & \textit{Describe}: Propose a classification framework to define the trolls’ identity and social role & US Elections
 \\
 
 \cite{Linvill2019a} & 3.2M tweets by 2K trolls & \textit{Describe}: Categorize trolls by role in political discussion & US Elections
 \\
 
 \cite{Llewellyn2019} & 3485 Brexit-related tweets by 419 trolls & \textit{Describe}: Analyze behavior shifts of trolls over time & Brexit
 \\
 
 \cite{Sobolev2019} & 500M posts by 700 trolls on Live Journal and 80K discussions with trolls’ participation & \textit{Assess the impact}: Identify the impact of trolls on conversations in social media & Political and social events in Russia
 \\
 
 \cite{Zannettou2019} & 1.8M images from 9M troll tweets & \textit{Describe}: Characterize image-posting activity of trolls & Broad, exploratory
 \\

 \cite{Zannettou2019a} & 27K tweets by 1K trolls; 
 
 96K tweets by 1K ordinary users & \textit{Describe; assess the impact}: Characterize trolls' operations and their influence on the greater Web & Broad, exploratory
 \\
 
\cite{Zannettou2019b} & 10M posts by 5.5K Russian (IRA) and Iranian trolls on Twitter and Reddit & \textit{Describe; assess the impact}: Identify and compare strategies of different groups of trolls (IRA vs Iranian), analyze the impact & Broad, exploratory 
\\
\bottomrule
\end{tabular}
\end{table*}

Quantitative troll studies can be divided into those (1) focusing on describing the behavior, strategy, and operations of trolls; (2) assessing the impact of trolls' operations; and (3) proposing methods for troll detection. The first group of studies are mostly descriptive and exploratory, characterizing the operations of trolls along multiple dimensions, including the use of hashtags, URLs, and retweets (e.g., \cite{gorrell2019partisanship, Zannettou2019a, Zannettou2019b}). The second group of studies typically assesses the impact of trolls' operations by tracking the engagement with trolls within the social network and in the greater Web \cite{Badawy2019, Sobolev2019, Zannettou2019a, Zannettou2019b}. The third group of studies often uses linguistic features (such as features common to native Russian speakers in English-language conversations) to detect trolls \cite{Boyd, Ghanem2019}, although some studies (e.g., \cite{Im2019}) use a broader set of features for troll detection, including profile-, behavior-, and stop word usage-related features. Quantitative troll studies could further be divided into those taking a broad scope, and those focusing on a particular propaganda campaign, most often the 2016 US Elections \cite{Kim2019, Linvill2019a, Howard2018, Addawood2019, Badawy2019, Boyd} and Brexit \cite{gorrell2019partisanship, Llewellyn2019}. This paper belongs to the first group of studies taking first a broad scope, after which focusing on the MH17 campaign.

Quantitative troll studies have typically analyzed textual data from Twitter (Table \ref{tab:quantitative_studies}). However, a few studies also examined Facebook ads purchased by IRA \cite{Boyd}, images posted on Twitter \cite{Zannettou2019}, as well as posts on other social networks, such as Reddit \cite{Zannettou2019b} and the blogging platform LiveJournal \cite{Sobolev2019} (popular in Russia). Furthermore, the first multi-platform troll studies have started to appear \cite{Howard2018}.

Previous research suggests that the behavior of IRA trolls significantly differs from ordinary social media users in posting patterns and language use. Namely, on Twitter, trolls often exhibit abnormal tweet and retweet rates, use more hashtags, and share more URLs \cite{Addawood2019, Im2019}. Moreover, they post shorter tweets with shorter words, use fewer words that indicate causation, and use less emojis \cite{Addawood2019, Boyd}. Furthermore, the trolls differ from each other. For example, \cite{Linvill2019a} identified five groups of trolls based on their behavior: right troll, left troll, news feed, hashtag gamer, and fearmonger. Finally, \cite{Kim2019} observed that the strategic behavior of trolls changes over time.

Among the studies of Russian trolls, our work most closely relates to \cite{Zannettou2019a, Zannettou2019b} in taking an exploratory approach and characterizing the activity of trolls across various dimensions. However, unlike \cite{Zannettou2019a, Zannettou2019b}, which analyze the content of English-language troll tweets and other troll activity (e.g., URL sharing) without language separation, we focus on Russian-language troll operations.

\subsection{The Crash of MH17}
Malaysia Airlines flight MH17 en route from Amsterdam to Kuala Lumpur was shot down in Eastern Ukraine on the 17th of July 2014. Shortly after the incident, western media accused separatists from the self-proclaimed Donetsk People's Republic (DPR) of shooting down the plane with Buk Missile System. In turn, the separatists and some Russian media stated that Ukrainian Armed Forces were to blame \cite{Oates2016}. The results of a criminal investigation by the Joint Investigation Team (JIT) published in September 2016 indicated that the Buk missile system was transported from Russia to territory controlled by separatists the day before the incident \cite{OpenbaarMinisterie2016}. 

The crash of MH17 sparked an intense debate on social media. In \cite{Golovchenko2018}, the authors collected tweets related to the incident and manually marked a sample of English-language tweets based on the judgements they expressed. 10.3\% of tweets appeared to be pro-Ukrainian, and 5.5\% -- pro-Russian, while the rest did not take a side. The authors found that pro-Russian and pro-Ukrainian tweets were mostly spread by accounts identified as ordinary citizens rather than media figures or politicians. However, they did not consider the role of trolls in the discussion. Furthermore, no academic studies have analyzed how trolls reacted to the crash of MH17.

Dutch journalists have analyzed the Twitter-released troll data and found that the two days after the crash of MH17 were the most active ever for the trolls in terms of the number of tweets \cite{VanderNoordaa2019}. 66000 of the July 18-19 tweets included the hashtags "Kiev Shot Down Boeing", "Kiev's Provocation", and "Kiev Tell the Truth". The authors further found that the first posts after the incident reported that the militia of DPR brought down a transport plane AN-26, although no plane other than MH17 was downed on that day.

\section{Content Analysis} \label{content_analysis}
\subsection{Hashtags and Hashtag Campaigns}
We first study the hashtags of the troll tweets. About 17\% of tweets include at least one hashtag (vs 47\%, $p<0.001$\footnote{Unless noted otherwise, the statistical tests in the paper are two-sided $X^2$ proportion tests from Stata 16.0 with clustered standard errors to account for, for example, multiple tweets per troll.}, for English-language troll tweets); with on average 1.39 hashtags per tweet with hashtags. Table \ref{tab:hashtags} shows the most used hashtags.

Similar to the previous work for English-language troll tweets \cite{Zannettou2019b}, we find that the most common Russian hashtag is \#News (\foreignlanguage{russian}{Новости}), which accounts for 4.96\% of all hashtags used by trolls (9.47\%, $p<0.01$, for hashtags used by English language trolls). Other popular apolitical hashtags include \#Auto (\foreignlanguage{russian}{Авто}), \#Sport (\foreignlanguage{russian}{Спорт}), and \#Cinema (\foreignlanguage{russian}{Кино}). Furthermore, several top hashtags relate to geographic locations, such as \#StPetersburg\footnote{More precisely, \textit{\#spb (\foreignlanguage{russian}{спб})}, which stands for Saint Petersburg}, \#Russia (\foreignlanguage{russian}{Россия}), and \#Ukraine (\foreignlanguage{russian}{Украина}). Some hashtags have a political sentiment, such as \#KievsProvocation (\foreignlanguage{russian}{ПровокацияКиева}), which was actively used after the crash of Malaysia Airlines' flight MH17. Other political hashtags relate to US (\#ReturnCalifornia -- \foreignlanguage{russian}{ВернитеКалифорнию}) and Ukraine (\#PanicInKiev -- \foreignlanguage{russian}{ПаникаВКиеве}). Interestingly, one of the top hashtags (\#RunZelensky -- \foreignlanguage{russian}{ЗеленскийБеги}) refers to Vladimir Zelensky, then a well-known figure in Russian media who was later elected as the president of Ukraine. Overall, we observe more event- and person-specific along with politically colored hashtags than in the 20 hashtags most used in English-language operations. 

\begin{table}[tbp!]
\caption{Trolls most used hashtags (translated)}
\label{tab:hashtags}
\begin{tabular}{p{2.25cm}p{0.9cm}p{2.25cm}p{0.9cm}}
\toprule
Hashtag & Share & Hashtag & Share \\
\midrule
News & 4.96\% & ImageOfRussia & 1.28\% \\ 
StPetersburg &  4.62\% & Putin & 1.08\% \\  
Russia & 3.45\% & Sport & 1.04\% \\
RussianSpirit & 2.65\% & ReturnCalifornia & 0.98\%\\ 
NevskieNews & 2.18\% & Cinema  & 0.94\% \\ 
KievsProvocation & 1.98\% & BattleOfOligarchs & 0.87\% \\
KievShot-DownBoeing & 1.97\% & Music & 0.80\% \\ 
KievTellTheTruth & 1.95\% & RunZelensky & 0.73\% \\
Ukraine & 1.85\% & Politics & 0.73\% \\ 
Auto & 1.30\% & Football & 0.72\% \\ 
\bottomrule
\end{tabular}%
\end{table}

\begin{figure}[tbp!]
\centering
\includegraphics[width=\columnwidth]{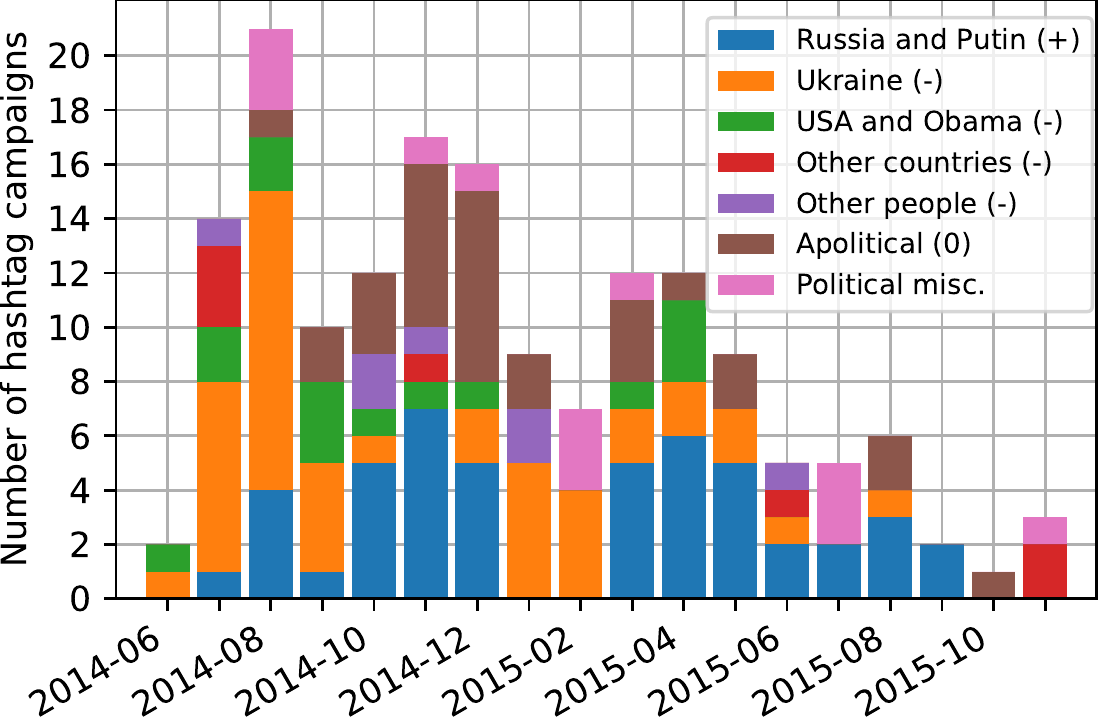}
\caption{Trolls' hashtag campaigns by category over time (+,-,0 indicates the positive, negative, or neutral sentiment of each campaign category)}
\Description{163 hashtag campaigns were detected. 48 praised Russia and Putin, 43 criticized Ukraine, and 15 criticized the US and Obama.}
\label{fig:hashtag_campaigns}
\end{figure}

Further, we check whether hashtags can indicate propaganda campaigns run by Russian trolls. To detect such propaganda \textit{hashtag campaigns}, we first select hashtags used by trolls more than 500 times (303 hashtags) with at least 95\% of their occurrences happened within one month (165 hashtags). Further, we explore tweets containing the selected hashtags and manually classify them based on their subject and sentiment. Some of the campaigns included several hashtags; we remove such duplicates (8 hashtags). Further, some hashtag campaigns had two subjects and sentiments, for example, attacking Ukraine and praising Russia simultaneously. In such cases, two subjects and sentiments are marked (7 campaigns). Figure \ref{fig:hashtag_campaigns} illustrates the focus of the trolls' hashtag campaigns over time. Most of the campaigns were run between June 2014 and November 2015, with only one detected outside this range (in April 2016, not shown in the figure). The analysis shows 163 campaigns divided into seven categories.

Figure \ref{fig:hashtag_campaigns} demonstrates that the main focuses of Russian-language trolls were praising Russia with Putin (48 campaigns), criticizing Ukraine (43 campaigns), as well as the USA with Obama (15 campaigns). We notice that about half of the Anti-Ukrainian and Anti-USA and Obama campaigns were run in July-September 2014. Anti-Ukrainian campaigns further exhibited an increase in January-February 2015, however, there was a surprising decline in October-December 2014. Patriotic pro-Russian and pro-Putin campaigns ran steadily between July 2014 and September 2015, except for January and February 2015. Eight campaigns criticized other countries: three were related to the European Union sanctions targeting Russia  (e.g., \#AgainstSanctions -- \foreignlanguage{russian}{ПротивСанкций}), two were against Canada (e.g., \#nocanada in November 2014), two attacked Turkey and were related to a Russian military jet downed by Turkey in Syria in November 2015 (e.g., \#TurkeyAggressor), and one criticized Armenia for demonstrations in June 2015 (\#YerevanBeSmart). Several campaigns attacked particular Russian and foreign figures (e.g., \#SomeoneWhoKillsChildren against the then president of Ukraine Petro Poroshenko). Further, about 18\% of all hashtag campaigns were apolitical. More than half of such hashtags appeared in October-December 2014, when anti-Ukraine campaigns were on the decline. Finally, among the miscellaneous hashtags with varying subjects and sentiments, three were pro-Ukraine campaigns run in February 2015 related to the Minsk II agreement\footnote{https://www.bbc.com/news/world-europe-31436513} (e.g., \#MinskHope -- \foreignlanguage{russian}{МинскаяНадежда}); two were anti-LGBT campaigns from July 2015 (e.g., \#StopLGBT -- \foreignlanguage{russian}{СтопЛГБТ}), and several campaigns addressed internal Russian events. For example, in December 2014 trolls blamed speculators for the Russian Ruble depreciation (\#Speculators -- \foreignlanguage{russian}{Спекулянты}). 

The hashtag campaign analysis illustrates the wide variation in topics that the trolls addressed. Interestingly, trolls devoted more attention to foreign rather than internal Russian affairs. Furthermore, in some campaigns, particularly anti-USA and -Obama, they addressed events that were not related to Russia. For example, trolls participated in the \#IHaveADream campaign about a black teen killed by a policeman in the US, and \#LatteSalute about Obama saluting Marines while holding a coffee cup. At the same time, only six campaigns criticized internal Russian events and figures. Also, although some of the hashtags were extensively used, only about 3\% of hashtags appeared in the data more than 100 times. Further, about 45\% of hashtags were used only once, as the blue curve in Figure \ref{fig:ECDF_Hashtags_RT_URL} illustrates, compared with 50\% ($p<0.001$) for English-language hashtags. Therefore, although trolls seem to have been somewhat free in choosing hashtags for their tweets, hashtag posting by Russian-language trolls was slightly more centralized than in English-language operations. 

\subsection{Retweets and Shared URLs}
Next, we analyze the use of retweets and URLs in troll tweets. Around 42.5\% of all tweets were retweets (RTs) (vs 44.2\%, $p>0.05$, for English-language troll tweets). While around 63.2\% contained a URL (vs 31.5\%, $p<0.001$, for English-language troll tweets and 26\% for ordinary user tweets \cite{Zannettou2019a}). Combined, RTs and tweets with a URL accounted for 75.6\% of all troll tweets. The portion of tweets with a RT or URL varied over time from around 20\% in February 2014 to 94\% in December 2015. Around 35.9\% of the retweets were the tweets of other trolls. Other retweets included posts of mass media accounts (with the top three being RIA Novosti, Gazeta.ru, and RT), public figures, and personal blogs.

\begin{figure}
\centering
\includegraphics[width=\columnwidth]{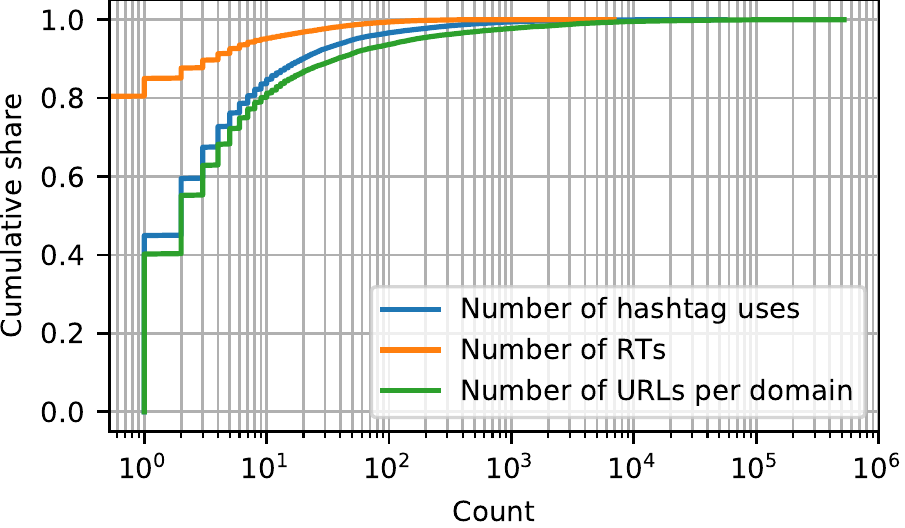}
\caption{ECDF of the counts of hashtag uses, retweets, and shared URLs per domain}
\Description{Only 20\% of tweets were retweeted at least once. About 45\% of hashtags were used once only.}
\label{fig:ECDF_Hashtags_RT_URL}
\end{figure}

Furthermore, we find that the average number of retweets of a trolls' tweet (with retweets of other trolls excluded) is 3.1. Only about 20\% of the original troll tweets were retweeted at least once, as the orange curve in Figure \ref{fig:ECDF_Hashtags_RT_URL} indicates. In addition, among retweeted tweets, the portion of retweeted tweets shared more than 100 times is only about 3\%. Therefore, engagement with trolls as measured by the number of retweets is relatively small.

\begin{figure}
\centering
\includegraphics[width=\columnwidth]{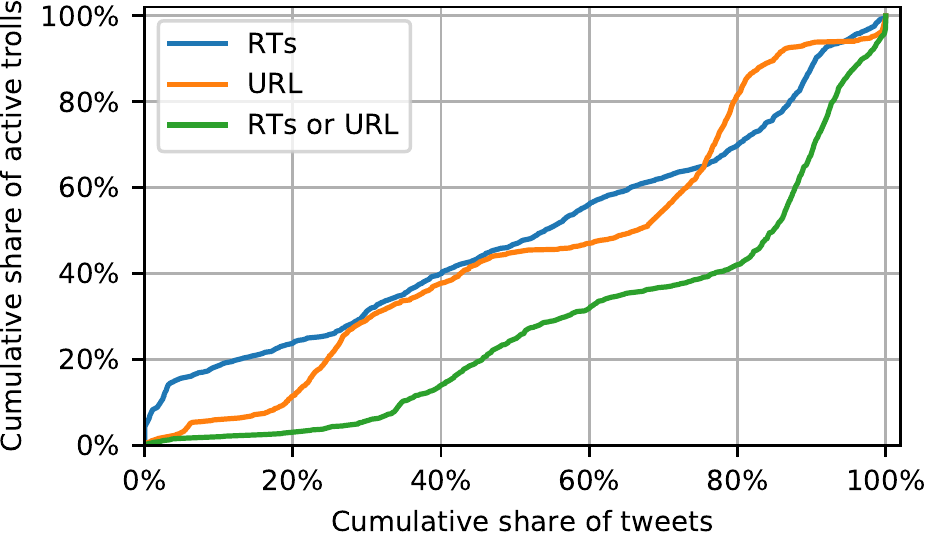}
\caption{ECDF of active trolls ($\geq$100 tweets/year) by tweet types}
\Description{Around 75\% of trolls re-shared content through RT and URLs in more than half of their tweets.}
\label{fig:accounts_by_tweet_type_ECDF}
\end{figure}

We further analyze retweeting and URL-sharing activity of trolls through the distribution of tweet types across ``active'' trolls that have at least 100 tweets in the data. Figure \ref{fig:accounts_by_tweet_type_ECDF} illustrates the empirical cumulative distribution functions (ECDF) of trolls by tweet types. The median share of RTs among trolls' tweets is around 54\%, meaning that RTs account for at least 54\% of tweets for half of the trolls. Furthermore, half of the trolls include a URL in 66\% of their tweets, which is much larger the share of 15\% reported for ordinary users \cite{Benevenuto2010}. This observation is consistent with the previous research \cite{Addawood2019}. Finally, 50\% of trolls included a RT or URL (or both) in 85\% of their tweets, thereby resharing information either directly (RT) or indirectly (URL). 

Furthermore, we analyze the most popular domains referred in troll tweets. Shortened links accounted for around 40\% of all shared URLs, dominated by the services bit.ly (62\% of shortened links), dlvr.it (11.1\%), and goo.gl (9.5\%). Therefore, we create a script that visits each link and captures the final domain after all redirections. However, due to unreachability of some shortened links, we capture the domain information of around 84\% URLs. Table \ref{tab:urls} shows the top 20 most common domains of URLs shared by trolls. More than one fifth of URLs link to a popular Russian blogging platform LiveJournal. Other frequently referred domains include the news agencies FAN (riafan.ru) and Neskiy News (nevnov.ru), which the US government links to IRA\footnote{https://home.treasury.gov/news/press-releases/sm577}; sites publishing news about Ukraine (kievsmi.net, kiev-news.com, and emaidan.com.ua), and social media platforms (youtube.com, vk.com, and twitter.com). We further analyze the number of URLs per domain and find that 40\% of domains are referred to only once, while 6\% of domains are referred to more than 100 times (as the green curve in Figure \ref{fig:ECDF_Hashtags_RT_URL} indicates). 

\begin{table}
\caption{The most common web domains referred to by trolls}
\label{tab:urls}
\begin{tabular}{p{2.1cm}p{1.15cm}p{2.1cm}p{1.15cm}}
\toprule
Domain & Share & Domain & Share\\
\midrule
livejournal.com & 20.7\% & inforeactor.ru & 1.5\% \\
riafan.ru & 16.8\% & vk.com & 1.5\% \\ 
gazeta.ru & 5.1\% & tass.ru & 1.3\% \\ 
ria.ru & 4.5\% & politexpert.net & 1.3\% \\ 
rt.com & 3.1\% & emaidan.com.ua & 1.1\% \\ 
nevnov.ru & 2.3\% & lenta.ru & 1.0\% \\ 
vesti.ru & 2.1\% & rbc.ru & 1.0\% \\
kievsmi.net & 1.8\% & twitter.com & 1.0\%\\  
youtube.com & 1.7\% & lifenews.ru & 0.9\%\\
kiev-news.com & 1.7\% & podrobnosti.biz & 0.8\%\\ 
\bottomrule
\end{tabular}%
\end{table}

\section{Temporal posting patterns} \label{temporal_patterns}

First, we analyze the temporal posting patterns of trolls. For each troll, we calculate a common temporal burstiness measure $B$ \cite{Goh2008} of their tweets' inter-arrival times (IATs, the times have a one minute granularity). The measure is defined as
\begin{displaymath}
  \ B= \frac{\sigma_{\tau} - \mu_{\tau}}{\sigma_{\tau} + \mu_{\tau}} 
\end{displaymath}
where $\mu_{\tau}$ is the sample mean and $\sigma_{\tau}$ is the sample standard deviation of the IAT distribution $\tau$. $B$ can vary from -1 to 1, with $B=1$ corresponding to a highly bursty signal, $B=0$ to a random (Poissonian) signal, and $B=-1$ to a highly regular (periodic) signal.

\begin{figure}
\centering
\includegraphics[width=\columnwidth]{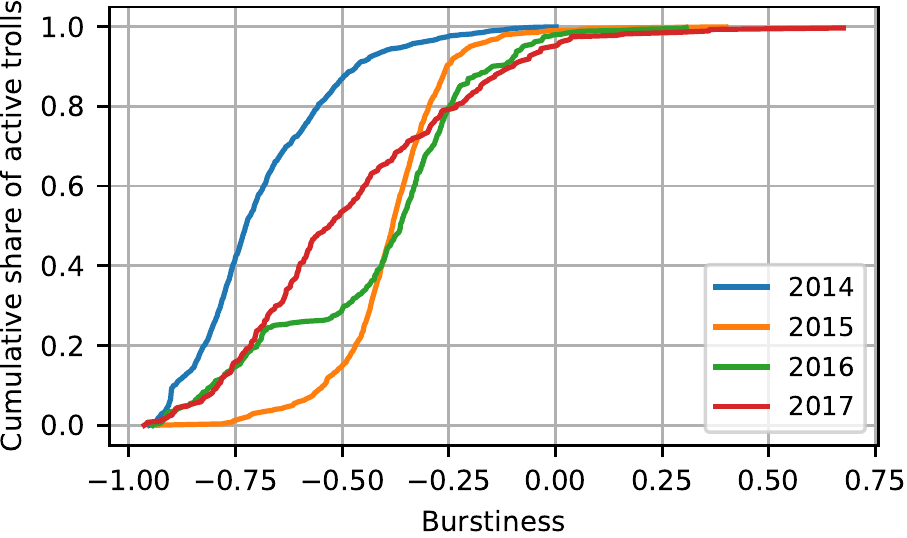}
\caption{ECDF of temporal burstiness of active trolls ($\geq$30 tweets/year) by year}
\Description{Burstiness is negative for most of the trolls suggesting the trolls exhibit period patterns.}
\label{fig:iat_burstiness}
\end{figure}

We calculate $B$ for each troll that posted at least 30 tweets in a given year. Figure \ref{fig:iat_burstiness} illustrates the ECDF of $B$ across trolls by year. We find that at least 95\% of trolls exhibited quite periodic posting patterns with $B<0$. The burstiness for ordinary twitter users (and for many other human driven activities) is around 0.2 to 0.4 \cite{Kim2016,Goh2008}. Therefore, we hypothesize that trolls use automation tools for tweet posting. We further observe that burstiness varied over the years. Namely, in 2014, posting patterns were the most periodic with a median $B=-0.73$, whereas in 2015 and 2016 the median burstiness increased to $-0.38$ and $-0.36$, respectively, potentially indicating a lower degree of the automation use. 

\begin{figure}
\centering
\includegraphics[width=\columnwidth]{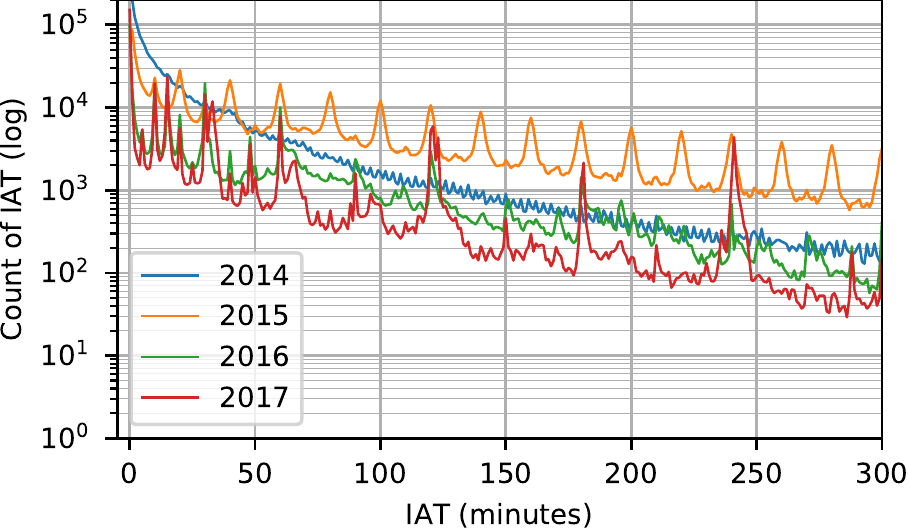}
\caption{Frequency distribution for the inter-arrival times of troll tweets by year}
\Description{Inter-arrival time (IAT) of trolls' tweets shows cyclical patterns that change over time.}
\label{fig:interarrival_time}
\end{figure}

Next, we analyze the periodicity of tweet posting on an aggregate level. Namely, we examine the frequency distribution for IATs that were calculated separately for each troll and pooled together. We construct the frequency distribution for every year from 2014 to 2017. The distributions in Figure \ref{fig:interarrival_time} show cyclical patterns in tweet posting activity. We observe that posting patterns also differ across years. Namely, in 2014, peaks of IATs were generally multiples of three, particularly, after IAT = 90. In 2015, peaks of IATs were spread around multiples of 20, generally in the interval $[20n - 2, 20n+1]$. In 2016-2017, patterns were similar, with IAT peaks at multiples of 30 minutes as well as at 10, 15, 20, and 48 minutes. However, in 2017 the hourly peaks were more prominent than in 2016.

Overall, the periodic temporal patterns in Figure \ref{fig:interarrival_time} reinforce the automation findings from Figure \ref{fig:iat_burstiness}. Furthermore, given that the analysis was conducted on an aggregate level, we infer that some trolls used automation tools with the same or similar settings. For example, in 2015, about 47\% of trolls had more IATs between $[20n - 2, 20n+1]$ (multiples of 20 min) than the 20\% of trolls that we would expect given uniformly random automation settings. Though we also note that the share of such automated tweets rarely exceeded 50\% for any troll suggesting trolls also likely manually posted or changed automation settings.

\section{Reacting to MH17 crash} \label{mh17}
Furthermore, we analyze the reaction of trolls to the crash of Malaysia Airlines flight MH17. 

\begin{figure}
\centering
\includegraphics[width=\columnwidth]{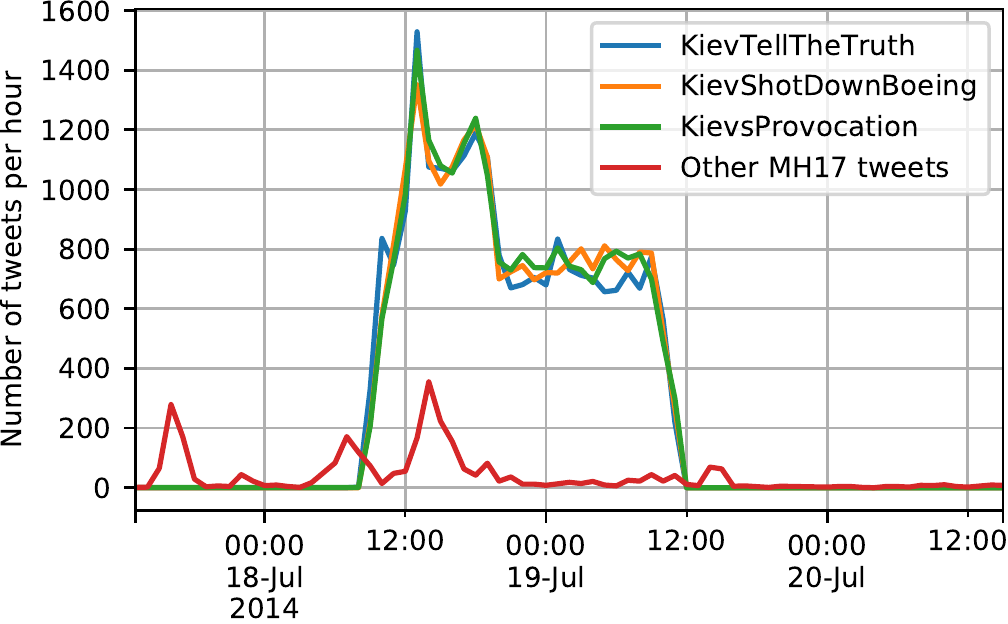}
\caption{Number of hourly troll tweets related to MH17 over time}
\Description{MH17 hashtag campaign started on 18.07.2014 at around 9:00 and ended abruptly at around 12:00 on 19.07.2014.}
\label{fig:MH17_tweet_count}
\end{figure}

Trolls used hashtags \#KievsProvocation (\foreignlanguage{russian}{ПровокацияКиева}), \#KievShotDownBoeing (\foreignlanguage{russian}{КиевСбилБоинг}), and \#KievTellTheTruth (\foreignlanguage{russian}{КиевСкажиПравду}) during the two days after the incident. We find troll tweets related to the incident using the above-defined hashtags and the keywords "mh17", "boeing" with Russian and English spellings, as well as Russian equivalents of the words "shot down" (\foreignlanguage{russian}{сбил*, сбит*}), "buk*" (\foreignlanguage{russian}{бук*}), "crash*" (\foreignlanguage{russian}{катастроф*}), "Malaysia*" (\foreignlanguage{russian}{малази*, малайз*}), and "plane" (\foreignlanguage{russian}{самолет*}). Furthermore, we only select tweets posted from 17.07.2014 14:00 (approximate time of the plane crash) until the end of July 2014. This results in 71023 tweets, 92.4\% of which were posted with at least one of the three mentioned hashtags, making the operation one of the largest run by the Russian trolls. This observation is in line with \cite{VanderNoordaa2019}. The content of tweets was often exactly the same for each of the three campaign hashtags. Nevertheless, perhaps due to space restrictions, most troll tweets used only one of the hashtags. Figure \ref{fig:MH17_tweet_count} shows the number of troll tweets related to the crash of MH17 published between 14:00 17.07 and 14:00 20.07, as 99\% of the such tweets fall into this interval. The hashtag campaigns started at around 9:00 on 18.07 and ended abruptly at around 12:00 on 19.07. The mean tweet rate during the campaign was 2484 tweets per hour, 9.4 times higher than the mean for July 2014. The mean hourly number of active trolls during the MH17 campaign was around 288, compared to 98 over all of July 2014.

\subsection{First reactions}
Table \ref{tab:first_MH17_tweets} shows the tweets that trolls posted during the first several hours after the crash including the tweet time (UTC) and the number of similar tweets posted within 10 minutes of each first tweet. This can show whether tweet posting was coordinated. 

The very first tweet related to the crash was posted at 15:01 stating that the militia of DPR had shot down a Ukrainian transport plane (an AN-26). The tweet referred to a news article published by the Federal News Agency\footnote{http://riafan.ru/2014/07/17/29576-opolchentsyi-dnr-sbili-transportnyiy-an-26-pod-gorodom-snezhnoe/}. The link to the article is still active, although the article has no connection to the original title indicated in the URL. At 15:40, the first tweets informing about the crash of MH17 appeared. At 15:59, the first tweet that blamed Ukraine for the incident was posted. Surprisingly, starting from 16:14, ten tweets appeared that repeated the statement that the DPR had shot down a Ukrainian transport plane but included the hashtag \#mh17. Moreover, one of the trolls first posted a tweet about the crash of MH17, and later about the shot-down transport plane. 

Finally, at 16:21 and 16:41 the authorities of DPR first denied their involvement and then blamed Ukrainian armed forces. At 16:45, a single tweet was posted stating that a plane was shot down by Russia with a link to a suspended blog post. At 17:19, some trolls posted tweets saying the DPR was framed by Ukrainian air controllers who had sent the plane into the firing zone. Finally, at 17:21 trolls stated, with the reference to the Luhansk People's Republic (LPR), that MH17 was shot down by a Ukrainian attack plane that was later destroyed.

The analysis of first reactions shows that at first trolls relied on the information from separatists that initially reported on downing of military AN-26 plane. Even after the crash of MH17 had been confirmed, some trolls did not seem to have realized that these two messages referred to the same incident. After the situation was clarified, trolls started to spread inconsistent messages placing responsibility on Ukraine, separatists, and even Russia. Therefore, it seems that in an initial {\it fog of war} scenario, the trolls were more interested in spreading confusion and mistrust rather than waiting to develop a coherent narrative and strategy. This method allows them to move quickly (within minutes of the incident) and in a decentralized manner.

\begin{table}
\caption{First reactions of trolls to MH17 crash and number of similar tweets within 10 mins}
\label{tab:first_MH17_tweets}
\begin{tabular}{p{0.7cm}p{5.3cm}p{1.3cm}}
\toprule
Time & Tweet text (translated) & Tweets in 10 min \\
\midrule
15:01 & The militia of DPR shot down a transport AN-26 near the town of Snizhne \textit{URL} & 14 \\ 
15:40 & Malaysian plane Amsterdam – Kuala Lumpur crashed on the border of Ukraine and Russia \textit{URL} & 19 \\
15:59 & Daily Bacchanalia in Ukraine. Today a plane was accidentally shot down. It’s terrible, comrades, how can one live like that? & 1 \\ 
16:14 & The militia reports: another plane of Ukrainian air forces was destroyed \#mh17 ukraine-russia  & 10 \\ 
16:21 & DPR denies the involvement in the crash of Malaysian plane  \textit{URL} & 10 \\
16:41 & The authorities of DPR blamed Ukrainian armed forces for the crash of Malaysian Boeing \textit{URL} & 19 \\
16:45 & A plane was shot down by Russia \textit{URL} & 1 \\ 
17:19 & RT: \#Ukraine frames (makes) \#DPR shoot down an international plane by sending it through air controllers to the firing zone \textit{URL} & 3 \\
17:21 & Boeing was shot down by Ukrainian attack plane, which was later destroyed - LPR (Luhansk People's Republic) \textit{URL} & 18 \\ 
\bottomrule
\end{tabular}%
\end{table}

\subsection{Tweet text analysis}
We observe that many tweets from different trolls include exactly the same text (while not being retweets). To test this, we tokenize tweets into sentences. We find 94 tokens that were used more than 500 times, with the most common being “why shoot down a civilian plane” (\foreignlanguage{russian}{зачем сбивать гражданский самолет}), “what did you expect” (\foreignlanguage{russian}{а вы чего ждали}), “do you agree with me” (\foreignlanguage{russian}{а вы со мной согласны}), and “here is what my friend posted” (\foreignlanguage{russian}{вот друг мой опубликовал}). Many tweets combined two or three of such sentences, e.g., \textit{"There are as many opinions as there are people. Why shoot down a civilian plane? Ukrops went totally nuts"}. Interestingly, some of these common sentences were also used in other anti-Ukrainian and anti-USA hashtag campaigns later on, such as \#ReturnCalifornia (\foreignlanguage{russian}{ВернитеКалифорнию}) in September 2014 and \#BattleOfOligarchs (\foreignlanguage{russian}{БитваОлигархов}) in March 2015.  

Further, we analyze the text of trolls' tweets related to the MH17 incident to understand their purpose and main message. Specifically we use a two-stage approach. In the first stage, we cluster tweets using a machine learning approach. In the second stage, we hand-code tweet clusters into six categories using an inductive content analysis approach \cite{Elo2008}.

\subsubsection{Clustering stage}
Tweets are first lemmatized using the Yandex MyStem 3.1 morphological analyzer\footnote{https://yandex.ru/dev/mystem/; https://github.com/nlpub/pymystem3}. Then stopwords and punctuation are removed. Apart from the common stopwords, we also remove the three most common hashtags discussed above. Further, tweets are tokenized into words, and small tweets with less than 7 tokens as well as tokens that occurred only once are filtered out. The remaining 41143 tweets contain 1401 unique tokens.

We calculate the term frequency-inverse document frequency (TF-IDF) for the corpus of tweets and the pairwise cosine similarities between the TF-IDF of tweets. We also test embedding-based representations of tweets, but TF-IDF performs well in grouping very similar (almost duplicate) tweets, and therefore we select it as a more intuitive approach. Next we cluster tweets using agglomerative hierarchical clustering. Average linkage is selected as it produces the most persuasive clusters. We test several cluster cut-off levels from 0.3 to 0.7 in increments of 0.05 and compare the resulting clusters using the silhouette coefficient. Lower cut-off levels result in a higher silhouette score but also larger numbers of clusters, particularly singletons. Therefore, we choose the cut-off level of 0.5 as a trade-off resulting in a reasonably small number of clusters for hand-coding (n = 873, including 200 singletons) and an acceptable silhouette coefficient (0.44)\footnote{At a cut-off level of 0.4, the silhouette score is only 7\% higher (0.47), yet the number of clusters is 115\% higher (1883).}.

\subsubsection{Coding stage}
The clusters are coded by the main author based on the text of five randomly selected tweets from each cluster. For clusters with less than 5 tweets, all available tweets are inspected. The coding scheme is developed iteratively to represent the main purpose and message of tweet clusters. To ensure the reliability of coding, 15\% of clusters (n=131) are also independently coded by another researcher with native Russian language skills. Cohen's kappa score of inter-rater reliability on the sample varies from 0.79 to 0.92 for different coding categories, indicating at least a substantial level of agreement \cite{Landis1977}.

For coding, first the tweet text (excluding hashtags) is examined to determine if the text actually mentions or implies the MH17 incident. If so, the tweet is also coded through the scheme described below. These directly related tweets are referred to as DR tweets. This initial filtering helps to remove hashtag-amplifying tweets with unrelated text.

The coding scheme for the DR tweets consists of the following non-exclusive categories:

\begin{enumerate}
\item \textit{Tweets blame Ukraine for the MH17 incident.} The tweets blame Ukraine either directly or indirectly for downing the plane or enabling the incident.

\item \textit{Tweets include the following narrative of blame on Ukraine.} For tweets in category (1), the narrative of blame is also grouped into several categories such as Ukrainians downed the plane or Ukrainians are responsible at least indirectly (e.g. for letting the incident happen).

\item \textit{Tweets contain news.} The tweets include information related to the incident that is presented as news rather than personal opinion and refers to a source (such as photos or reports of eyewitness) or is based on expert opinions.

\item \textit{Tweets contain disinformation or propaganda.} For tweets in category (3), the tweets contains disinformation (i.e., contradicts the findings of JIT) or presents information in a biased way to support a particular narrative\footnote{News containing erroneous information, which does not suggest a responsible party or an alternative version of the incident, is not considered as disinformation.} (propaganda news).

\item \textit{Tweets contain the following narrative of disinformation and propaganda.} For tweets in category (4), the narrative of disinformation and propaganda is grouped into four high-level and 16 low-level categories as shown in Table \ref{tab:fake_news_narratives}.
\end{enumerate}

\subsubsection{Findings}
We find that 68\% of the troll tweets (when hashtags are excluded) do not actually mention or imply the incident (i.e., non-DR tweets). Therefore, these tweets seemingly served only as hashtag-amplifiers. Furthermore, 99.9\% of such tweets were published during the hashtag campaign (between 9:00 18.07 and 12:00 19.07) compared to 89.9\% for DR tweets, supporting the hypothesis of hashtag amplification. Nevertheless, almost all of such non-DR tweets discuss Ukraine in a more general context, typically in a negative tone.

Surprisingly, tweets blaming Ukraine for the incident account for only 49\% of the DR tweets. Nevertheless, this share is much larger than 5.5\% of pro-Russian tweets detected by \cite{Golovchenko2018} for English-language tweets related to MH17. Table \ref{tab:narratives_of_blame} shows the distribution for the narratives of blame among such tweets. About 73\% of such tweets suggest that Ukraine downed the plane and 21\% say that Ukraine is responsible at least indirectly (e.g., by allowing the plane to fly over the war zone or by escalating the conflict in Ukraine). Further, 3\% blame Ukraine for interfering with the crash investigation, mainly by bombing the crash site. Finally, around 3\% point to the potential responsibility of Ukraine without directly blaming, e.g., by recalling a similar incident with a Russian Tu-155 accidentally shot down by the Ukrainian Air Force in 2001 or stating that at the time of the crash MH17 was in a Ukrainian air-defense zone.

\begin{table}
\caption{The narratives of blame placed on Ukraine for the crash of MH17 of troll tweets}
\label{tab:narratives_of_blame}
\begin{tabular}{p{2.3cm}p{4cm}p{1cm}}
\toprule
Narrative & Example tweet & Share \\
\midrule
Ukraine downed the plane & News of Ukraine. According to an expert, the Boeing 777 was shot down by Ukrainian air defense with 90\% probability & 73\% \\ 
Ukraine is responsible at least indirectly & Vladimir Putin: Ukraine is responsible for the Boeing crash & 21\% \\
Ukraine interferes with the investigation & Latest news: Ukraine does not allow Malaysian experts to the Boeing wreckage & 3\% \\ 
Ukraine may be responsible & Just read it. Experts recall Tu-155 shot down 13 years ago & 3\% \\ 
\bottomrule
\end{tabular}%
\end{table}

Furthermore, about 57\% of the DR tweets contain news. Most of the news tweets inform about the number of crash victims, their nationality, the progress of evidence collection, or report the statements of officials on the incident. However, 23\% of news tweets (13\% of all DR tweets) contain disinformation or propaganda. Table \ref{tab:fake_news_narratives} shows the main narratives of such tweets. About half of the disinformation suggests that a Ukrainian fighter aircraft downed the plane. According to trolls, the size of debris indicated that the plane was downed by an air-to-air missile. Trolls further referred to a statement from Spanish air-traffic controller's working in Kiev, which claimed to have seen a military aircraft escorting MH17\footnote{This statement was later disproven as Ukrainian laws require air traffic controllers to have Ukrainian citizenship}. About 23.1\% of disinformation tweets suggest that Ukraine downed the plane from the ground, with 11.8\% not openly blaming but stating that Ukraine relocated their Buk missile systems to the location of the missile launch a day before the crash. Interestingly, two contradicting disinformation campaigns (Ukrainians shot down the plane by an air-to-air vs. ground-to-air missile) were run in parallel. Overall, trolls' tweets seem to reflect the changing and contradicting narratives of the Russian government on MH17 \cite{Toler2018}.

Furthermore, 21.4\% of disinformation and propaganda tweets use other narratives to blame Ukraine. Trolls suggested that the incident was a planned operation, and that the President of Ukraine knew about the crash before it happened because his reaction to the crash was so fast. Finally, about 5.6\% of disinformation tweets do not blame Ukraine, including a few posts where trolls confused MH17 for a military AN-26 (see Table \ref{tab:first_MH17_tweets}) and a few tweets with experts supporting conspiracy theories. 

\begin{table}
\caption{The narratives of disinformation and propaganda of troll tweets}
\label{tab:fake_news_narratives}
\begin{tabular}{p{6.3cm}p{1cm}}
\toprule
Narrative & Share \\
\midrule
\textit{Ukrainian fighter aircraft downed the plane:} & \textit{49.8\%}\\ 
- Debris are too large & 24.2\% \\
- Statement of air controller & 10.2\% \\ 
- Eyewitness report & 10.1\% \\ 
- Using machine guns & 2.1\% \\ 
- Other & 3.2\% \\ 
\midrule

\textit{Ukrainians downed the plane from the ground:} & \textit{23.1\%} \\ 
- Ukrainians relocated Buks shortly before the incident & 11.8\% \\
- Assassination attempt on Putin & 7.1\% \\
- Using air defense systems (general) & 2.9\% \\ 
- Mistake at military training & 1.3\% \\ 
\midrule

\textit{Other narratives (blaming Ukraine):} & \textit{21.4\%} \\ 
- Ukraine interferes with the investigation & 7.4\% \\ 
- President of Ukraine knew about the incident before it happened & 7.1\% \\
- Traffic controllers purposefully diverted the plane to the war zone & 1.7\% \\ 
- Miscellaneous & 5.2\% \\ 
\midrule

\textit{Other narratives (not blaming Ukraine):} & \textit{5.6\%} \\ 
- Separatists shot down AN-26 & 1.3\% \\ 
- Conspiracy theories & 0.5\% \\ 
- Miscellaneous & 3.8\% \\ 
\bottomrule
\end{tabular}%
\end{table}

\section{Discussion and conclusion} \label{discussion}
This paper shed light on domestic and regional Russian-language operations of Russian trolls on Twitter by analyzing trolls' tweets from between 2014 and 2017. Namely, we first studied the tweets' content through their hashtags, the use of retweets, and shared URLs. Further, we analyzed tweets' temporal posting patterns. After that, we focused on trolls' reaction to Malaysia Airlines MH17 crash. 

\textbf{Tweet Content.} Comparing the results of the hashtag analysis with English-language hashtags of Russian trolls from a previous study \cite{Zannettou2019b}, we see that in both cases trolls did not focus solely on political and social issues, but also used apolitical hashtags (e.g., \#Sport) likely to appear similar to ordinary users. Nevertheless, at least eight out of the top 20 hashtags in Russian were related to politics, compared with three for English-language operations \cite{Zannettou2019b}. This difference is mainly due to the ``hashtag campaigns'' that Russian-language trolls often ran, which accounted for six of the eight political hashtags, and which did not seem to be common in English-language operations. A potential motivation for such campaigns is to increase the hashtag visibility in the country-specific Twitter trends. Therefore, a hashtag campaign is more sensible in the smaller Russian domain of Twitter rather than a larger English language domain. 

We discovered 165 hashtag campaigns, which were run between June 2014 and November 2015. Their main sentiments were attacking Ukraine, USA and Obama personally, as well as praising Russia and Putin. The targets of such campaigns stayed roughly the same over the 18 months. However, about half of anti-Ukraine and anti-USA campaigns were run between July-September 2014, when the international pressure on Russia increased following the crash of MH17. Further, no campaigns praising Russia and Putin appeared in January-February 2015. This could be related to the depreciation of Russian ruble that peaked during these months. Moreover, there was also no anti-USA campaigns during the months.

We further found that trolls actively reshared information by using retweets (RTs) and URLs. The share of original tweets (i.e., without RTs or URLs) was only about 24\%, and only about a quarter of trolls posted original tweets more often than tweets with RTs or URLs. About two thirds of trolls' retweets were the posts of non-troll accounts. Though at least half of the top 20 most retweeted non-troll accounts and most referenced internet domains were of mainstream media. However, an average troll tweet received only about 3.1 retweets from ordinary users, showing that the engagement with trolls’ posts was relatively low. 

\textbf{Temporal posting patterns.}
We analyzed the burstiness and frequency distribution of tweet inter-arrival times (IATs) to understand their temporal posting patterns. Using burstiness we discovered that, unlike normal Twitter users, many trolls exhibited highly periodic posting patterns. Furthermore, the frequency distribution of pooled IATs revealed three distinct cyclical patterns that prevailed during different years. Namely, in 2014, peaks in IATs were detected at multiples of three minutes, in 2015, multiples of 20 minutes, and in 2016-2017, multiples of 30 minutes. Such patterns could indicate the use of tweet posting automation tools with similar settings. Therefore, Russian trolls could more accurately be described as cyborgs or bot-assisted humans \cite{Chu2012}.

\textbf{Reacting to MH17 crash.}
In reaction to the crash of Malaysia Airlines flight MH17, Russian-language trolls ran their largest single information campaign (by the number of tweets). However, 68\% of the 71K tweets of the campaign had text not directly related to the incident; such tweets were seemingly used only for hashtag amplification. Nearly half of the remaining (related) tweets blamed Ukraine for the crash, either alleging that Ukraine downed the plane (73\%), or suggesting some degree of responsibility (21\%). Surprisingly, only 13\% of such related tweets contained news-like disinformation or propaganda. Furthermore, the narratives of such fake news were not internally consistent. Namely, approximately half stated that Ukraine shot down the plane with an air-to-air missile, whereas about 23\% suggested the use of a ground-to-air missile. The fake news also reported diverse reasons for the incident, ranging from framing the separatists and Russia, to a mistake in military training, to a failed assassination attempt on Putin. 

\begin{acks}
The authors thank Anar Bazarhanova for helping with coding the tweets. Benjamin Finley is supported by the 5GEAR project (No. 319669) and the FIT project (No. 325570) both funded by the Academy of Finland.
\end{acks}

\bibliographystyle{ACM-Reference-Format}
\bibliography{paper}

\end{document}